\documentclass[prb,aps,twocolumn,floats,showpacs,amsmath,amssymb]{revtex4}

\usepackage{graphics}
\usepackage{amsmath}
\usepackage{dcolumn}

\begin{document}

\title{
{\rm\small\hfill (submitted to Phys. Rev. B)}\\
CO oxidation on Pd(100) at technologically relevant pressure conditions:\\
A first-principles kinetic Monte Carlo study}

\author{Jutta Rogal\cite{address}}
\author{Karsten Reuter}
\author{Matthias Scheffler}
\affiliation{Fritz-Haber-Institut der Max-Planck-Gesellschaft,
Faradayweg 4-6, D-14195 Berlin, Germany}

\received{30th January 2008}

\begin{abstract}
The possible importance of oxide formation for the catalytic activity of transition metals in heterogenous oxidation catalysis has evoked a lively discussion over the recent years. On the more noble transition metals (like Pd, Pt or Ag) the low stability of the common bulk oxides suggests primarily sub-nanometer thin oxide films, so-called surface oxides, as potential candidates that may be stabilized under gas phase conditions representative of technological oxidation catalysis. We address this issue for the Pd(100) model catalyst surface with first-principles kinetic Monte Carlo (kMC) simulations that assess the stability of the well-characterized $(\sqrt{5} \times \sqrt{5})R27^{\circ}$ surface oxide during steady-state CO oxidation. Our results show that at ambient pressure conditions the surface oxide is stabilized at the surface up to CO:O$_2$ partial pressure ratios just around the catalytically most relevant stoichiometric feeds ($p_{\rm CO}:p_{\rm O_2} = 2:1$). The precise value depends sensitively on temperature, so that both local pressure and temperature fluctuations may induce a continuous formation and decomposition of oxidic phases during steady-state operation under ambient stoichiometric conditions.
\end{abstract}

\pacs{82.65.+r, 68.43.Bc, 68.43.De, 68.35.Md}


\maketitle

\section{Introduction}
The large difference in catalytic activity of ``ruthenium'' catalysts towards CO oxidation under ultra-high vacuum and ambient pressure conditions \cite{peden86} has recently been controversially discussed in terms of a possible oxide formation at the surface in technologically relevant environments \cite{over03,reuter06a}. The well known much lower thermodynamic stability of the bulk oxides of Pd or Pt \cite{reuter04} suggests at first glance that this discussion has no direct relevance for these more noble metals that are equally, if not more often used in oxidation catalysis.  Using a constrained first-principles thermodynamics approach \cite{reuter03} we could show in a preceding study \cite{rogal07}, however, that the situation is not that clear-cut when considering for the possibility of sub-nanometer thin oxide films, so-called surface oxides. Mapping out a wide range of $(T,p_{\rm O2},p_{\rm CO})$-conditions we specifically found for the Pd(100) surface that the stability range of the well-characterized $(\sqrt{5} \times \sqrt{5})R27^{\circ}$ ($\sqrt{5}$ in brevity) surface oxide structure \cite{orent82,chang88a,chang88b,zheng02,lundgren04} does indeed extend up to pressure and temperature conditions relevant for technological CO oxidation catalysis $(T=300-600$\,K, $p_{\mathrm{O}_2} = p_{\mathrm{CO}} \sim 1$\,atm).

This insight motivates to center a detailed kinetic modeling of the surface structure and composition during steady-state CO oxidation at technologically relevant pressure conditions on the $\sqrt{5}$ surface oxide. More precisely, the question of a possible oxide formation at the surface can be narrowed down to addressing the stability, formation and decomposition of the $\sqrt{5}$ surface oxide structure in the reactive environment. Theoretically, one is then still faced with the challenge to evaluate the kinetics of the surface elementary processes over time scales of the order of seconds or longer, while accurately accounting for the evolving atomistic structure of the surface. In this study, we meet the time-scale challenge with a first-principles statistical mechanics approach, combining density-functional theory (DFT) for an accurate description of the kinetic parameters with kinetic Monte Carlo (kMC) simulations for the evaluation of the non-equilibrium statistical mechanics problem. At present, such an approach is preferentially carried out on the basis of a lattice model for the surface in order to keep the number of required DFT-based parameters tractable. Due to the different periodicity and Pd-density of the $\sqrt{5}$ overlayer ~\cite{todorova03,rogal07,kostelnik07} and the Pd(100) substrate, a combined lattice model embracing both structures will be highly involved and will require numerous kinetic parameters for a manifold of conceivable atomistic pathways for the complex structural transition between the oxidized and pristine Pd surface. Before embarking on such an explicit modeling of the formation and decomposition of the surface oxide structure in the simulations it is thus appropriate to first assess whether the stability range of the $\sqrt{5}$ surface oxide does indeed extend to the catalytically relevant environments, when the kinetics of the ongoing CO$_2$ formation is fully accounted for. In this study we determine this stability range by focusing on the onset of the decomposition of the surface oxide structure when the partial pressure ratio $p_{\rm CO}:p_{\rm O_2}$ exceeds a critical value. We identify this onset once the average O concentration in the $\sqrt{5}$ structure is reduced below the value corresponding to the intact surface oxide, which allows us to restrict our kMC modeling to the $\sqrt{5}$ lattice.

The central result of our kMC simulations, which we had briefly highlighted before \cite{rogal07a}, is that at ambient pressures the surface oxide is stabilized at the surface up to CO:O$_2$ partial pressure ratios just around the catalytically most relevant stoichiometric feeds ($p_{\rm CO}:p_{\rm O_2} = 2:1$). This suggests an interpretation of the morphological changes observed in recent reactor scanning-tunneling microscopy (STM) measurements \cite{hendriksen04} under O-rich feeds in terms of a formation of the $\sqrt{5}$ structure at the surface. The precise value for the critical partial pressure ratio we compute depends furthermore sensitively on temperature, so that both local pressure and temperature fluctuations may induce a continuous formation and decomposition of the oxidic phase during steady-state operation under ambient stoichiometric conditions. We therefore expect that a full understanding of the catalytic activity of the Pd(100) surface in such technologically relevant environments can only be obtained from a modeling that explicitly accounts for the corresponding on-going structural changes, which will then also be able to address the actual role played by the surface oxide for the catalytic activity itself. Nevertheless the rather high peak turnover frequencies obtained in our simulations even at the intact surface oxide show already that the formation of this structure in the reactive environment can not simplistically be equated with a ``deactivation'' of the model catalyst.

\section{Theory}
\label{sec:theory}
\subsection{First-principles kinetic Monte Carlo simulations}
Our target is to describe the steady-state catalytic CO oxidation over the $\sqrt{5}$ surface oxide by explicitly accounting for the kinetics of the different elementary processes taking place in the system. Properly evaluating the time evolution of such a system requires simulation cells that are large enough to capture the effects of correlation and spatial distribution of the chemicals at the surface. Furthermore, most of the considered processes are highly activated and occur on time scales that are orders of magnitude longer than e.g. a typical vibration ($\sim 10^{-12}$\,s). Due to these so-called rare events we need to evaluate the statistical interplay between the different elementary processes over an extended time scale that can reach up to seconds or more. Since this is unfeasible for molecular dynamics simulations, we employ kinetic Monte Carlo simulations, where the time evolution is efficiently coarse-grained to the rare-event dynamics. In a kMC simulation, each kMC-step corresponds to the execution of a rare event, while appropriately accounting for the short-time dynamics of the system in between. In the generated sequence of system states, each state is assumed to be independent of all previous states, i.e. the kMC algorithm provides an efficient numerical solution to the Master equation of a continuous time Markov process~\cite{bortz75,gillespie76,voter86,kang89,fichthorn91}.

In order to keep the number of considered elementary processes in a tractable range the kMC simulations are performed on a lattice.  Obviously, the chosen lattice model has to properly reflect the important structural features of the real system.  For this model a list of \emph{all relevant} elementary processes $p$ on the lattice is set up with the corresponding rate constants $k_p$ that characterize the probability to escape from the present system state by the process $p$.  Starting in a given initial system configuration the sum over the rate constants of all \emph{possible} processes in this current configuration is evaluated, $k_{\mathrm{tot}} = \sum_p k_p$. One process $i$ is then randomly chosen by
\begin{eqnarray}
\label{eq:ratesum}
\sum_{p=0}^{i-1} k_p < \rho_1 k_{\mathrm{tot}} \leq \sum_{p=0}^{i} k_p \quad ,
\end{eqnarray}
with $\rho_1 \in ]0,1]$ being a random number.  Since the probability of selecting a process $i$ is weighted by its rate constant $k_i$ a process with a large rate constant is more likely to be chosen than a process with a small rate constant.  The selected process is executed and the system configuration is updated.
This way the kMC algorithm simulates a sequence of Poisson processes and an explicit relationship between kMC time and real time can be established~\cite{fichthorn91}.
This evolution of real time after each kMC step is given by
\begin{eqnarray}
\label{eq:kmctime}
t \rightarrow t - \frac{\ln(\rho_2)}{k_{\mathrm{tot}}} \quad ,
\end{eqnarray}
where $\rho_2 \in ]0,1]$ is a second random number~\cite{fichthorn91}.

The crucial ingredients to a meaningful kMC simulation are therefore the identification of all relevant elementary processes $p$ and the determination of the corresponding rate constants $k_p$. For the latter we rely on the modern first-principles kinetic Monte Carlo approach~\cite{ruggerone97,ovesson99,hansen00a,hansen00b,fichthorn00,kratzer02,reuter04b,reuter06} and calculate the rate constants using DFT and transition-state theory (TST). Concerning the elementary processes during the CO oxidation over the $\sqrt{5}$ surface oxide we consider adsorption, desorption, diffusion and reaction of the chemicals, and follow the recipe of how to determine the rate constants for these processes from DFT and TST given in Ref.~\onlinecite{reuter06}. In the next Section the resulting expressions for the rate constants are briefly reviewed.
The first-principles data required to evaluate the rate constants with these expressions is then given in Section~\ref{subsec:parameters}.

\subsection{The rate constants}
\label{subsec:rates}
\subsubsection{Adsorption}
The rate constant of adsorption of a gas phase species $i$ on a surface site of type $st$ depends on the impingement rate and the local sticking coefficient, $\tilde{S}_{i,st}$~\cite{reuter06}
\begin{eqnarray}
\label{eq:rateads}
k_{i,st}^{\mathrm{ads}} (T,p_i) =
\tilde{S}_{i,st} (T) \frac{p_i A}{\sqrt{2 \pi m_i k_{\mathrm{B}} T}} \quad .
\end{eqnarray}
The only system-specific parameters entering the impingement rate are the mass $m_i$ of the impinging gas phase molecules and the surface area $A$. An explicit, accurate evaluation of the local sticking coefficient $\tilde{S}_{i,st}$  is highly involved and requires a determination of the full, high-dimensional potential energy surface (PES) of the adsorption process, on which molecular dynamics (MD) simulations have to be performed for a statistically relevant number of trajectories~\cite{gross98}.
In the present study the local sticking coefficient is instead roughly estimated as~\cite{reuter06}
\begin{eqnarray}
\label{eq:sticking}
\tilde{S}_{i,st}(T) =
f_{i,st}^{\mathrm{ads}}(T)
\left( \frac{A_{i,st}}{A} \right)
\exp \left(- \frac{\Delta E_{i,st}^{\mathrm{ads}}}{k_{\mathrm{B}} T} \right) \quad ,
\end{eqnarray}
where $\Delta E_{i,st}^{\mathrm{ads}}$ is the highest barrier along the minimum energy pathway (MEP) and $A_{i,st}$ is the so-called active area. Within the concept of an active area it is assumed that only particles of a species $i$ with an initial lateral position within a certain area $A_{i,st}$ around the adsorption site $st$ can actually stick to this site. The rate constant of adsorption in Eq.~(\ref{eq:rateads}) will then actually become independent of the total impingement area $A$ and only the active area $A_{i,st}$ has to be known. $f_{i,st}^{\mathrm{ads}}$ is a factor that reduces the number of impinging gas phase particles by the fraction that is not traveling along the MEP and might thus be reflected at some higher barrier along a different pathway.

\subsubsection{Desorption}
Since the desorption process is the time-reversed process of adsorption the desorption rate constants are connected to the  adsorption rate constants by the detailed balance criterion and can thus be obtained by applying~\cite{reuter06}
\begin{eqnarray}
\label{eq:ratedes}
k^{\mathrm{des}}_{i,st}& = &
k^{\mathrm{ads}}_{i,st} \exp\left(-\frac{\Delta G_{i,st}(T,p_i)}{k_{\mathrm{B}} T}\right) \\ \nonumber
& \approx & k^{\mathrm{ads}}_{i,st} \frac{1}{z^{\mathrm{vib}}_{i,st}}
\exp\left(-\frac{\Delta \mu_{i,\mathrm{gas}}(T,p_i) - E^{\mathrm{bind}}_{i,st}}{k_{\mathrm{B}} T}\right)
\quad ,
\end{eqnarray}
where $\Delta G_{i,st}(T,p_i)$ is the change in the Gibbs free energy between a particle in the gas phase and in the adsorbed state.
$\Delta G_{i,st}(T,p_i)$ is approximated by the difference between the chemical potential of the particle in the gas phase, $\mu_{i,\mathrm{gas}}(T,p_i) = E^{\mathrm{tot}}_{i,\mathrm{gas}} + \Delta \mu_{i,\mathrm{gas}}(T,p_i) $ and the free energy of the particle in the adsorbed state $F_{i,st} = E^{\mathrm{tot}}_{i,st} - k_{\mathrm{B}}T \ln(z^{\mathrm{vib}}_{i,st})$.
To evaluate the desorption rate constant using Eq.~(\ref{eq:ratedes}) we thus need to determine, in addition to the adsorption rate constant,  the vibrational partition function of the particle in the adsorbed state, $z^{\mathrm{vib}}_{i,st}$, the temperature and pressure dependent part of the gas phase chemical potential, $\Delta \mu_{i,\mathrm{gas}}(T,p_i)$, as well as the binding energy, $E^{\mathrm{bind}}_{i,st} = E^{\mathrm{tot}}_{i,\mathrm{gas}}-E^{\mathrm{tot}}_{i,st}$ of the particle at its adsorption site $st$.

\subsubsection{Diffusion}
In a diffusion process a particle $i$ moves from a site $st$ to a site $st'$.  If an appropriate saddle point along the diffusion pathway can be identified, harmonic TST can be applied to obtain the corresponding rate constant of diffusion~\cite{reuter06}
\begin{small}
\begin{equation}
\label{eq:ratediff}
k^{\mathrm{diff}}_{i,st \to st'} (T) = f^{\mathrm{diff,TST}}_{i,st \to st'}(T) \left(\frac{k_{\mathrm{B}} T}{h}\right) \exp\left(-\frac{\Delta E^{\mathrm{diff}}_{i,st \to st'}}{k_{\mathrm{B}} T} \right)
\end{equation}
with
\begin{equation}
\label{eq:ratediff1}
 f^{\mathrm{diff,TST}}_{i,st \to st'}(T) = \frac{z^{\mathrm{vib}}_{i,st \to st', \mathrm{TS}}}{z^{\mathrm{vib}}_{i,st}}
\end{equation}
and
\begin{equation}
\label{eq:diffbarrier}
\Delta E^{\mathrm{diff}}_{i,st \to st'} = E^{\mathrm{tot}}_{i,st \to st', \mathrm{TS}} - E^{\mathrm{tot}}_{i,st}
\quad .
\end{equation}
\end{small}
The diffusion barrier $\Delta E^{\mathrm{diff}}_{i,st \to st'}$ denotes the maximum barrier along the MEP of the diffusion process and is given by the energy difference of the transition state ($E^{\mathrm{tot}}_{i,st \to st', \mathrm{TS}}$) and the initial state ($E^{\mathrm{tot}}_{i,st}$) at site $st$. The reverse process of the diffusion $st \to st'$ is simply the backward diffusion $st' \to st$ and therefore the rate constants of these two processes have to fulfill the detailed balance criterion. To calculate the diffusion rate constants the factor $f^{\mathrm{diff,TST}}_{i,st \to st'}$ (given by the ratio of the vibrational partition functions in the transition state (TS), $z^{\mathrm{vib}}_{i,st \to st', \mathrm{TS}}$, and the initial state $z^{\mathrm{vib}}_{i,st}$)   as well as the diffusion barriers have to be known.

\subsubsection{Reaction}
If a suitable reaction coordinate can be identified that contains a saddle point, again TST can be used to determine the reaction rate constant. In a general form  the reaction rate constant can then be written as~\cite{reuter06}
\begin{eqnarray}
\label{eq:ratereac}
\lefteqn{k^{\mathrm{reac}}_{i \to f} (T) =}\\ \nonumber
& & = f^{\mathrm{reac, TST}}_{i \to f} (T)
\left(\frac{k_{\mathrm{B}} T}{h}\right)
\exp\left(-\frac{\Delta E^{\mathrm{reac}}_{i \to f}}{k_{\mathrm{B}} T}\right)
\, ,
\end{eqnarray}
with $i$ denoting the initial state and $f$ the final state of the reaction process.
As for the diffusion events the factor $f^{\mathrm{reac, TST}}_{i \to f}$ can be obtained from the ratio of the partition functions in the transition and initial state. In addition the reaction barrier, $\Delta E^{\mathrm{reac}}_{i \to f}$, defined as the energy difference in the transition and initial state, has to be determined.

\subsection{Computational Setup}
\label{sec:comp}
The most crucial input parameters to determine the rate constants as discussed in the previous Section are the energy barriers for adsorption, diffusion and reaction, as well as the binding energies at a certain adsorption site. We calculate the total energies that are needed to evaluate these quantities using DFT and within the full-potential (linearized) augmented plane wave + local orbital (L)APW+lo method~\cite{sjoestedt00,madsen01} as implemented in the \textsf{WIEN2k} code~\cite{wien2k}.
The exchange-correlation (xc) energy is treated within the generalized gradient approximation (GGA) using the PBE~\cite{perdew96} xc-functional. All surfaces are simulated using the supercell approach with inversion symmetric slabs consisting of five Pd(100) layers with the reconstructed surface oxide layer plus additional O/CO on both sides. We verified that further increasing the number of Pd(100) layers to seven changes the binding energies of both adsorbates by an insignificant amount ($\leq 2$\,meV per adsorbate). The vacuum between consecutive slabs is at least 14\,\AA.  The outermost adsorbate layers, the surface oxide and topmost palladium substrate layer are fully relaxed. All calculations of gas-phase molecules (O, CO, CO$_2$) are done in rectangular supercells with side lengths of $(13 \times 14 \times 15)$\,bohr (for O), $(13 \times 14 \times 18)$\,bohr (for O$_2$ and CO) and $(13 \times 14 \times 20)$\,bohr (for CO$_2$). 

The muffin-tin radii are set to $R^{\mathrm{Pd}}_{\mathrm{MT}}=2.0$\,bohr for palladium, $R^{\mathrm{O}}_{\mathrm{MT}}=1.0$\,bohr for oxygen and $R^{\mathrm{C}}_{\mathrm{MT}}=1.0$\,bohr for carbon. Inside the muffin-tins the wave functions are expanded up to $l^{\mathrm{wf}}_{\mathrm{max}} = 12$ and the potential up to $l^{\mathrm{pot}}_{\mathrm{max}} = 6$. For the $\sqrt{5}$ surface unit cell a $[4 \times 4 \times 1]$ Monkhorst-Pack (MP) grid (8 irreducible k-points) has been used to integrate the Brillouin zone (BZ), and for larger surface unit cells the MP-grid has been reduced accordingly to assure an equivalent sampling of the BZ. For the gas-phase molecule calculations in the large rectangular supercells $\Gamma$-point sampling was employed. The energy cutoff for the expansion of the wave function in the interstitial is $E^{\mathrm{wf}}_{\mathrm{max}} = 20$\,Ry and for the potential $E^{\mathrm{pot}}_{\mathrm{max}} = 196$\,Ry.
The chosen computational setup is thus exactly the same as in the preceding thermodynamic study on this system~\cite{rogal07}.
As detailed there~\cite{rogal07} using these basis set parameters the reported binding energies per oxygen atom or per CO molecule are converged to within 50\,meV, which is fully sufficient for the arguments and conclusions presented here.

\subsection{The kMC model}
\label{subsec:kmcmodel}
Having established the theoretical framework for the kMC simulations the next crucial step is to build a suitable lattice model that represents the investigated system. For this, we use the insight gained in our preceding constrained thermodynamics study~\cite{rogal07} that identified the $\sqrt{5}$ as most relevant oxidic structure under catalytically interesting gas phase conditions ($p_i \sim 1$\,bar, $T \sim 300-600$\,K). As initially discussed in Section I we thus concentrate the kMC simulations on this particular structure and develop a first-principles kMC lattice model that allows for a proper mapping of the $\sqrt{5}$ surface oxide structure.

\subsubsection{The kMC lattice}
\begin{figure}
\scalebox{0.5}{\includegraphics{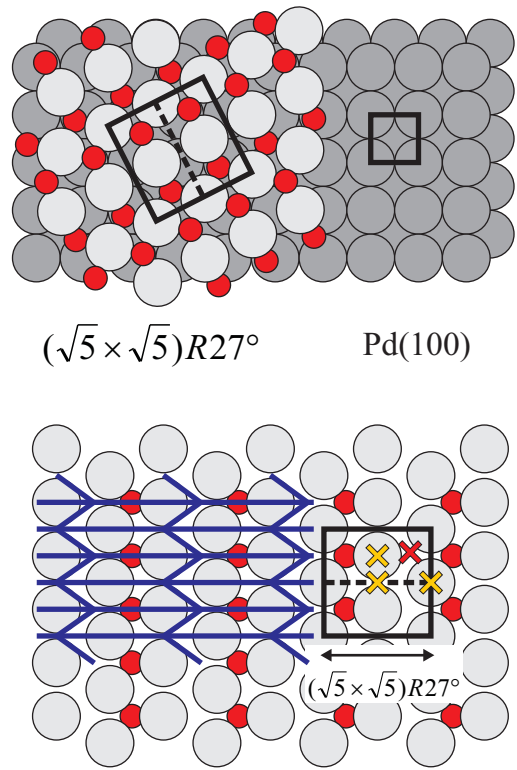}}
\caption{\label{fig1}
(Color online) In the upper panel a schematic top view of the $\sqrt{5}$ surface oxide structure is shown.
The large, grey spheres represent Pd atoms, the small, red (dark) ones O atoms.  The Pd atoms of the Pd(100) substrate are darkened. The $\sqrt{5}$ surface unit cell, as well as the $(1 \times 1)$ surface unit cell of Pd(100) are indicated.
As illustrated the $\sqrt{5}$ surface unit-cell consists of two halves that differ only with respect to the underlying Pd(100) substrate.  We find the difference in binding energies at the equivalent adsorption sites in the two halves to be very small, and correspondingly use the reduced unit-cell in the kMC simulations. In the lower panel these prominent adsorption sites and the final kMC lattice are illustrated. The three high-symmetry adsorption sites (bridge, top2f, top4f) are indicated by the yellow (light) crosses.  The hollow adsorption site of the upper O atoms (red (dark) cross) is likewise included as a site in the kMC lattice, whereas the reconstructed Pd layer as well as the lower O atoms are fixed.  In the final kMC lattice only the bridge and hollow sites are considered as shown on the lower left (see text).}
\end{figure}

In the upper panel of Fig.~\ref{fig1} a schematic illustration  of the surface oxide on Pd(100) is shown.
The $\sqrt{5}$ surface oxide structure characterized in UHV corresponds essentially to a (O-Pd-O)-trilayer of PdO(101) on top of the Pd(100) substrate~\cite{todorova03}. Each $\sqrt{5}$ surface unit-cell contains four Pd atoms in the reconstructed oxide layer. In this stoichiometric termination of the surface oxide two of the Pd atoms are fourfold coordinated and two are twofold coordinated by oxygen. There are also two kinds of oxygen atoms, two of them sit on-top of the reconstructed Pd layer (upper O atoms) and two at the interface to the Pd(100) substrate (lower O atoms) forming the previously mentioned trilayer structure.

At first sight the surface oxide structure exhibits six high-symmetry adsorption sites within the $\sqrt{5}$ surface unit cell, where either oxygen or CO can be adsorbed.~\cite{rogal07} These are two bridge sites (br), two twofold by oxygen coordinated top sites (top2f) and two fourfold by oxygen coordinated top sites (top4f). The difference between each of these doubly existing adsorption sites arises only from the arrangement of the underlying Pd(100) lattice.  These small structural deviations seem to have little consequences on the binding at these different prominent adsorption sites. The computed binding energies of O and CO in the two bridge sites differ by less than 0.1\,eV. Similarly for the two top2f and top4f sites the difference is even smaller, $\Delta E^{\mathrm{bind}} < 50$\,meV (a more detailed discussion of the binding energetics can be found in Ref.~\onlinecite{rogal07}). In our kMC model, we therefore neglect these small differences and build the lattice from the half sized $\sqrt{5}$ surface unit-cell as shown in Fig.~\ref{fig1}, each containing three different adsorption sites, bridge, twofold top and fourfold top.

Inspecting the computed energetics at these three different adsorption sites in more detail the kMC lattice model can even be further simplified.  We find that in the top2f and top4f sites the CO is bound much weaker, i.e. by $0.3-0.8$\,eV, compared to the bridge site.  In addition, the two top sites correspond only to very shallow PES minima, with diffusion barriers of less than 50\,meV to a neighboring bridge site.  Considering further that due to the computed strongly repulsive interactions~\cite{rogal07} it is not possible to adsorb CO at a top site next to an already occupied bridge site, we realize that the metastable top sites will effectively never be occupied. For oxygen the situation is even more pronounced, since here the additional adsorption of an O atom at the $\sqrt{5}$ surface oxide is only possible in the bridge site, whereas the two top sites are not even metastable. The two different top sites are therefore disregarded in setting up the kMC lattice representation of the surface oxide.

On the other hand, considering only the adsorption into prominent high-symmetry sites at the perfect $\sqrt{5}$ structure is not sufficient to address the CO-induced decomposition of this lattice.  For this, we assume that the surface oxide layer starts to destabilize by the removal of the upper oxygen atoms from the trilayer structure.
We therefore also include the corresponding hollow sites that are completely occupied by these surface oxygen atoms in the perfectly stoichiometric $\sqrt{5}$ structure as explicit sites in the kMC model, cf. Fig.~\ref{fig1}.
The basis for the final kMC lattice model is thus formed by the fixed reconstructed Pd layer and the lower O atom in the PdO trilayer.  On this lattice there are two different active site types (br and hol, cf. left part of the lower panel in Fig.~\ref{fig1}). Each site can either be vacant, or filled with one oxygen atom, or one CO molecule, whereas multiple adsorption at a site is not possible.

Considering all non-concerted processes possible on this lattice we obtain a total number of 26 possible processes.
This includes five adsorption processes, the unimolecular adsorption of CO into br or hol sites and the dissociative adsorption of O$_2$ into adjacent br--br, br--hol or hol--hol sites, as well as the corresponding five desorption processes.
Furthermore there are eight site and element specific diffusion processes describing the diffusion of O/CO from br$\to$br, hol$\to$hol, br$\to$hol and hol$\to$br sites, as well as four reaction processes including the reaction of O$^{\textrm{br}}$ + CO$^{\textrm{br}}$,  O$^{\textrm{hol}}$ + CO$^{\textrm{hol}}$, O$^{\textrm{br}}$ + CO$^{\textrm{hol}}$ and O$^{\textrm{hol}}$ + CO$^{\textrm{br}}$. Since the reaction processes are  modeled as associative desorption there are an additional four processes for the dissociative adsorption of CO$_2$. This list of non-concerted processes is exhaustive with respect to the chosen kMC lattice setup.  However, it can not be excluded that there might be also other, more complex, concerted processes which could have an influence on the here discussed system.

\subsubsection{Simulation setup}
All kMC simulations are performed on the final lattice model containing $(10 \times 10)$ half $\sqrt{5}$ surface unit cells, i.e. 100 br and 100 hol sites, with periodic boundary conditions.  To ensure that the size of the chosen lattice is sufficient simulations were also performed on larger lattices containing up to 1250 br and hollow sites, respectively, without observing significant differences concerning the investigated quantities.
These quantities of interest for the present study are
the average surface occupations $\overline{\theta}_{i,st}$ of a particle $i$ on a site $st$ under steady-state conditions, which is evaluated as
\begin{equation}
\label{eq:avocc}
\overline{\theta}_{i,st} = \frac{\sum_n \theta_{i,st,n} \cdot \Delta t_n}{\sum_n \Delta t_n} \qquad ,
\end{equation}
where $\Delta t_n$  is the time step of the $n$th Monte Carlo move.
In all simulations we checked carefully using different starting configurations and varying simulation lengths that the system actually reached steady-state, before the average surface occupations were evaluated.

\subsection{Evaluation of the rate constants}
\label{subsec:parameters}
For all 26 processes that are possible in the kMC model, the rate constants have to be determined.
The respective expressions for the four different process types, i.e. adsorption, desorption, diffusion and reaction, have been described in Section~\ref{subsec:rates}.
Here we determine the corresponding parameters entering the equations for the different rate constants, following the procedure described in more detail in Ref.~\onlinecite{reuter06}.

\subsubsection{Adsorption}
To calculate the rate constants of adsorption in Eq.~(\ref{eq:rateads})  we specify the active area $A_{i,st}$ for the impingement of each species on each site as simply half the surface area of the surface unit cell in the kMC lattice.  As shown in Fig.~\ref{fig1} this area corresponds to half the surface unit cell of the $\sqrt{5}$ structure ($1/2 A_{\sqrt{5}} = 19.47$\AA$^2$)
i.e. $A_{\mathrm{CO,br}} = A_{\mathrm{CO,hol}} = 1/4 A_{\sqrt{5}}$ for the adsorption of CO and $A_{\mathrm{O}_2,\mathrm{br-br}} = A_{\mathrm{O}_2,\mathrm{hol-hol}} = A_{\mathrm{O}_2,\mathrm{br-hol}} = 1/6 A_{\sqrt{5}}$ for the adsorption of O$_2$.

To obtain an estimate for the local sticking coefficient $\tilde{S}_{i,st}$ in Eq.~(\ref{eq:sticking})
we perform DFT calculations to determine first of all if there are any significant barriers, $\Delta E^{\mathrm{ads}}_{i,st}$, along the MEP for any of the five different adsorption processes.
For the CO molecule this was done by vertically lifting the CO in an upright geometry from a bridge resp. hollow site. For different heights of the CO molecule above its adsorption site the energy of the whole system was then minimized with respect to all other degrees of freedom .  For neither of the two sites a barrier was found along this desorption resp. adsorption pathway, i.e. $\Delta E_{\mathrm{CO,br}}^{\mathrm{ads}} = \Delta E_{\mathrm{CO,hol}}^{\mathrm{ads}} = 0.0$\,eV.
For the dissociative adsorption of O$_2$ molecules a PES was mapped out in which the energy is given as a function of the bond length between the two O atoms and the height above the surface of their center of mass (so-called elbow-plot).  The lateral position of the center of mass within the surface unit cell, as well as the orientation of the O$_2$ molecule were kept fixed.  Since there are only two hollow sites within the $\sqrt{5}$ surface unit cell a lifting of two oxygen atoms would correspond to a complete removal of all upper oxygen atoms in the surface oxide structure due to the periodic supercell setup of the DFT calculations. Therefore, for these calculations the size of the  $\sqrt{5}$ surface unit cell was doubled in the direction of the involved sites. We find that the dissociation of an O$_2$ molecule in this particular orientation appears to be non-activated. A similar result is also found for the O$_2$ adsorption in neighboring bridge-hollow sites, i.e. $\Delta E_{\mathrm{O}_2,\mathrm{hol-hol}}^{\mathrm{ads}} = \Delta E_{\mathrm{O}_2,\mathrm{br-hol}}^{\mathrm{ads}} = 0.0$\,eV.  For the adsorption in two neighboring bridge sites, though, the PES reveals a high barrier of  $\Delta E_{\mathrm{O}_2,\mathrm{br-br}}^{\mathrm{ads}} = 1.9 $\,eV.

A second parameter entering the sticking coefficient in Eq.~(\ref{eq:sticking}) is
 the factor $f_{i,st}^{\mathrm{ads}}$ representing the reduction in sticking due to the fraction of impinging molecules not following the MEP.  Since the kMC simulations are performed in a temperature range of $T = 300 - 600$\,K, where the impinging gas phase molecules can still be considered as rather slow, it is assumed that the molecules are quite efficiently steered along the MEP  and thus the value of $f_{i,st}^{\mathrm{ads}}$ is approximated by $f_{i,st}^{\mathrm{ads}} \approx 1$ for all five adsorption processes. With these parameters the sticking coefficients are unity for the four barrier-free adsorption processes.  For the activated adsorption of O$_2$ in two neighboring bridge sites the sticking coefficient is as small as $\sim 10^{-16}$ even for a temperature of $T = 600$\,K. Even if the crude way of extracting the approximate activation barrier from the restricted elbow plot would thus introduce an error of many orders of magnitude in the sticking coefficient, this adsorption process will therefore still be insignificant compared to the adsorption of O$_2$ in two neighboring hollow or neighboring bridge and hollow sites. The results of the kMC simulations reported below where indeed found to be unaffected, even when lowering the activation barrier for this adsorption process by up to 0.5\,eV.

\subsubsection{Desorption}
Having determined the five different adsorption rate constants the respective desorption rate constants can be calculated using detailed balance via Eq.~(\ref{eq:ratedes}). To a first approximation the vibrational partition function in the adsorbed state entering Eq.~(\ref{eq:ratedes}) is set to unity, $z^{\mathrm{vib}}_{i,st} \approx 1$, for all five desorption processes. In the gas phase this approximation would certainly be reasonable over the here investigated temperature range ($T = 300 - 600$\,K), i.e. only the first vibrational state of the O$_2$ and CO molecules is excited.  In the adsorbed state, though, this might be changed due to the lower frequency modes resulting from the adsorbate-substrate bond, which could then lead to a somewhat larger partition function, $z^{\mathrm{vib}}_{i,st} > 1$, and therefore slightly smaller desorption rate constants. Since this enters only linearly into the desorption rate constant we nevertheless found that considering such increased partition functions did not significantly affect the conclusions drawn from the simulations below.

The temperature and pressure dependent part of the gas phase chemical potential $\Delta \mu_{i,\mathrm{gas}}(T,p_i)$ is calculated  by approximating the O$_2$ and CO gas phase with ideal gas phase reservoirs and evaluating the translational, vibrational, rotational, electronic and nuclear partition functions using statistical thermodynamics~\cite{rogal06}.  This approximation by an ideal gas only introduces a negligible error in the temperature and pressure range considered here~\cite{rogal06}.

\begin{figure}
\scalebox{0.5}{\includegraphics{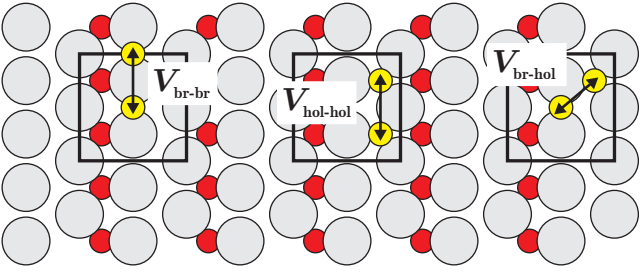}}
\caption{\label{fig2}
(Color online) Considered nearest neighbor pair interactions between adsorbates on the $\sqrt{5}$ surface oxide lattice.  Large, grey spheres represent Pd atoms. Small, red spheres oxygen atoms included in the kMC lattice and small yellow spheres adsorbed O and/or CO.}
\end{figure}

A further parameter that is needed to evaluate Eq.~(\ref{eq:ratedes}) is the binding energy  of the particles in their respective adsorption sites, $E^{\mathrm{bind}}_{i,st}$.  Since the binding energies of O and CO in bridge and hollow sites also depend on lateral interactions between the adsorbates, we expand them in a lattice gas Hamiltonian (LGH)~\cite{stampfl99,reuter05,fontaine94,sanchez84,zunger94}. Aiming only at a first approximation of the effect of lateral interactions we employ a crude LGH that is restricted to the nearest neighbor pair interactions between adsorbates in different and alike sites, yielding
\begin{eqnarray}
\label{eq:lgh}
H &=&
\sum_i \left[n_{\mathrm{O}, i} E_{\mathrm{O}, i}^{\,0} + n_{\mathrm{CO}, i} E_{\mathrm{CO}, i}^{\,0}\right] \\ \nonumber
& & + \sum_{ij} [V_{\mathrm{O-O}, ij} \, n_{\mathrm{O}, i} \, n_{\mathrm{O}, j}
+  V_{\mathrm{CO-CO}, ij} \, n_{\mathrm{CO}, i} \, n_{\mathrm{CO}, j} \\ \nonumber
& & \quad \quad +  V_{\mathrm{O-CO}, ij} \, n_{\mathrm{O}, i} \, n_{\mathrm{CO}, j}]
\quad ,
\end{eqnarray}
where $i$ runs over all lattice sites in the kMC lattice and $j$ only over the corresponding nearest neighboring sites.  $E_{\mathrm{O}, i}^{\,0}$ and $E_{\mathrm{CO}, i}^{\,0}$ are the on-site energies of O and CO on a site $i$, $V_{i,j}$ is the interaction parameter between two adsorbates sitting on sites $i$ and $j$ and $n_i$ is the occupation number, i.e. $n_i = 0$ denotes that site $i$ is empty and $n_i = 1$ that it is occupied.
To obtain an estimate of the error introduced by limiting the LGH to the nearest neighbor pair interactions only we performed  DFT calculations with sparser adsorbate arrangements
in larger surface unit cells, i.e. in $(2\sqrt{5} \times \sqrt{5})R 27^{\circ}$ and $(\sqrt{5} \times 2\sqrt{5})R 27^{\circ}$ cells.
The obtained binding energies of O and CO in these larger surface unit cells differ by less than 0.1\,eV from the respective values calculated within the $\sqrt{5}$ surface unit cell.  This change in binding energies is within the here assumed level of accuracy, rationalizing the choice to neglect lateral interactions between adsorbates extending beyond the $\sqrt{5}$ surface unit cell in the current study. Employing the LGH in Eq.~(\ref{eq:lgh}) the binding energy of a CO molecule in a site $i$ is then given by
\begin{eqnarray}
\label{eq:lgh_binding}
\lefteqn{E_{\mathrm{CO},i}^{\mathrm{bind}} = } \\ \nonumber
&=& H(n_{\mathrm{CO},i} = 1) - H(n_{\mathrm{CO},i} = 0) \\ \nonumber
&=& E_{\mathrm{CO}, i}^{\,0} + 2 \sum_j \left[V_{\mathrm{CO-CO}, ij} \, n_{\mathrm{CO}, j}
+ V_{\mathrm{O-CO}, ij} \, n_{\mathrm{O}, j} \right]
\quad .
\end{eqnarray}
Accordingly, the binding energy of an oxygen atom is calculated.  During the associative desorption of an O$_2$ molecule, respectively, two neighboring sites are depleted, which can be expressed likewise using the LGH.  As shown in Fig.~\ref{fig2} the first nearest neighbor pair interactions correspond to interactions between two neighboring bridge sites, two neighboring hollow sites and between neighboring bridge and hollow sites.
This generates six different interaction parameters between like species ($V_{\mathrm{O-O, br-br}}$, $V_{\mathrm{O-O, hol-hol}}$, $V_{\mathrm{O-O, br-hol}}$, $V_{\mathrm{CO-CO, br-br}}$, $V_{\mathrm{CO-CO, hol-hol}}$, $V_{\mathrm{CO-CO, br-hol}}$) and four different ones between unlike species ($V_{\mathrm{O-CO, br-br}}$, $V_{\mathrm{O-CO, hol-hol}}$, $V_{\mathrm{O-CO, br-hol}}$, $V_{\mathrm{O-CO, hol-br}}$).  Together with the four on-site energies we thus have to determine 14 parameters for the LGH.  This is achieved by fitting the parameters to DFT total energies of 29 different ordered configurations with O and/or CO in bridge and hollow sites within the $\sqrt{5}$ surface unit cell, and expressing the respective energy of each configuration in terms of the LGH expansion (cf. Eq.~(\ref{eq:lgh})).  Since within the kMC model the lower oxygen atoms of the $\sqrt{5}$ surface oxide trilayer form a part of the lattice itself the energies of the different configurations are calculated with respect to the \emph{empty} kMC lattice, i.e.
\begin{small}
\begin{eqnarray}
\label{eq:lgh_ene}
\lefteqn{\Delta E(\mathrm{O,CO@}\sqrt{5} - 2 \mathrm{O})} \\ \nonumber
& & = E_{\mathrm{O,CO@}\sqrt{5}- 2 \mathrm{O}} - E_{\sqrt{5} - 2 \mathrm{O}}
- N_{\mathrm{O}} 1/2 E^{\mathrm{tot}}_{\mathrm{O}_2} - N_{\mathrm{CO}} E^{\mathrm{tot}}_{\mathrm{CO}} \quad ,
\end{eqnarray}
\end{small}
where $E_{\sqrt{5} - 2 \mathrm{O}}$ corresponds to the total energy of the surface oxide structure without the two upper oxygen atoms.
The total energy of an oxygen molecule, $E^{\mathrm{tot}}_{\mathrm{O}_2}$, is calculated  using  the relationship $E^{\mathrm{tot}}_{\mathrm{O}_2} = 2 E^{\mathrm{tot}}_{\mathrm{O}} + E^{\mathrm{bind}}_{\mathrm{O}_2}$ with a highly converged value of the oxygen gas phase binding energy of $E^{\mathrm{bind}}_{\mathrm{O}_2} = -6.22$\,eV~\cite{kiejna06}.  Table~\ref{tab:lgh_ene} lists the determined values for the four on-site energies and 10 interaction parameters.
\begin{table}
\caption{\label{tab:lgh_ene}
Four on-site energies $E^{\,0}$ and 10 first nearest neighbor pair interaction parameters $V$ for the lattice gas Hamiltonian describing the adsorption of O and CO in hollow and bridge sites on the $\sqrt{5}$ surface oxide.  All values in eV.}
\begin{ruledtabular}
\begin{tabular}{llll}
\rule[0mm]{0mm}{4mm}$E_{\mathrm{O, br}}^{\,0}$  &   $E_{\mathrm{O, hol}}^{\,0}$  &   $E_{\mathrm{CO, br}}^{\,0}$  &   $E_{\mathrm{CO,hol}}^{\,0}$\\[0.5ex]
                    -0.51       &   -1.95   &   -1.40   &   -1.92\\[1ex]
\hline
\rule[0mm]{0mm}{4mm}$V_{\mathrm{O-O, br-br}}$    &   $V_{\mathrm{O-O, hol-hol}}$  &   $V_{\mathrm{O-O, br-hol}}$\\[0.5ex]
                    0.08        &   0.07    &   0.08    \\[1ex]
\hline
\rule[0mm]{0mm}{4mm}$V_{\mathrm{CO-CO, br-br}}$    &   $V_{\mathrm{CO-CO, hol-hol}}$  &   $V_{\mathrm{CO-CO, br-hol}}$\\[0.5ex]
                    0.08        &   0.13    &   0.14    \\[1ex]
\hline
\rule[0mm]{0mm}{4mm}$V_{\mathrm{O-CO, br-br}}$    &   $V_{\mathrm{O-CO, hol-hol}}$  &   $V_{\mathrm{O-CO, br-hol}}$ &   $V_{\mathrm{O-CO, hol-br}}$\\[0.5ex]
                    0.06        &   0.11    &   0.13    &   0.12    \\[1ex]
\end{tabular}
\end{ruledtabular}
\end{table}
With these parameters  the binding energy of every adsorbate for any random configuration on the lattice can be evaluated using Eq.~(\ref{eq:lgh_binding}).  With this binding energy the respective desorption rate constants can then be calculated via Eq.~(\ref{eq:ratedes}).

\subsubsection{Diffusion}
Within the kMC lattice model for the $\sqrt{5}$ surface oxide there are eight different diffusion processes, four between like sites (O/CO hol$\to$hol and br$\to$br) and four between unlike sites (O/CO hol$\to$br and br$\to$hol).
To obtain the corresponding rate constants of diffusion the vibrational partition functions of the adsorbates in the initial state and in the transition state are needed, as well as the diffusion barriers, cf. Eqs.~(\ref{eq:ratediff}) and~(\ref{eq:ratediff1}).

\begin{figure}
\scalebox{0.25}{\includegraphics{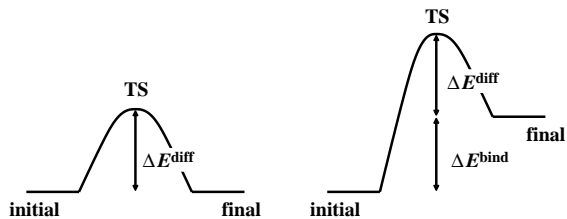}}
\caption{\label{fig3}
Schematic illustration of the diffusion barriers between energetically like and unlike sites.}
\end{figure}

With a similar justification as for the desorption rate constants, we assume the partition functions in initial and transition state to be comparable, i.e. $z^{\mathrm{vib}}_{i,st} \approx z^{\mathrm{vib}}_{i,st \to st', \mathrm{TS}}$ yielding $f^{\mathrm{diff,TST}}_{i,st \to st'} \approx 1$ for all eight diffusion processes. To calculate the diffusion barriers, $\Delta E^{\mathrm{diff}}_{i,st \to st'}$, we determine the transition state by defining a reaction coordinate connecting the initial and final state and calculate the energy of O/CO for different positions along this coordinate.  Since we found the surface oxide trilayer to be rather mobile with respect to a registry shift parallel to the Pd(100) substrate~\cite{rogal07}, the lateral positions of the reconstructed Pd atoms are additionally kept fixed in these calculations. The resulting barriers for these approximate diffusion paths are listed in Table~\ref{tab:diffbarrier}.
\begin{table}
\caption{\label{tab:diffbarrier}
Barriers for the diffusion between the different sites within the $\sqrt{5}$ surface unit cell.  All values are in eV.}
\begin{ruledtabular}
\begin{tabular}{lccc}
\rule[0mm]{0mm}{4mm}
&  $\Delta E^{\mathrm{diff}}_{\mathrm{br}\to\mathrm{br}}$  &   $\Delta E^{\mathrm{diff}}_{\mathrm{hol}\to\mathrm{hol}}$  &   $\Delta E^{\mathrm{diff}}_{\mathrm{hol}\to\mathrm{br}}$\\[0.5ex]
\hline
\rule[0mm]{0mm}{4mm}O                       &    1.2         &   1.4      &       0.1    \\
CO                      &    0.4        &   0.6      &       0.3  \\[0.5ex]
\end{tabular}
\end{ruledtabular}
\end{table}
The barriers have been calculated within the $\sqrt{5}$ surface unit cell with both hollow sites occupied by O (O/CO diffusion br $\to$ br), only one hollow site occupied by O (O/CO diffusion hol $\to$ br) and one hollow site occupied by either O or CO (O/CO diffusion hol $\to$ hol). The initial and the final states of a diffusion process can either be energetically degenerate (left part of Fig.~\ref{fig3}) or -- for the diffusion between unlike sites or like sites with a different nearest neighbor coordination -- the energies of the initial and final state wells differ (right part of Fig.~\ref{fig3}).
In the latter case the values listed in Table~\ref{tab:diffbarrier} represent the minimum value of the diffusion barrier, i.e. if the final state has the same or a lower energy than the initial state the diffusion barrier is equal to the tabulated value.  If on the other hand the final state is higher in energy the difference in binding energies between the initial and final state is added to the listed barrier to ensure that the detailed balance criterion is fulfilled for the two time-reversed diffusion processes. Since this difference in binding energies depends on the nearest neighbor coordination of the two lattice sites involved in the diffusion process, lateral interaction effects are this way also considered in the diffusion process rate constants.

The described approximations in determining the exact transition states lead to uncertainties in the stated barrier values which in turn influence the corresponding diffusion rate constants. We checked on the importance of the various diffusion processes on the simulation results and found them to play only a minor role. Essentially, even completely switching off all diffusion processes led only to insignificant changes of the results presented below. We are thus confident that the rather coarse procedure to determine the diffusion barriers is sufficient with respect to the current purpose of the simulations.

\subsubsection{Reaction}
To calculate the rate constants for the four possible reaction processes (O$^{\mathrm{br}}$+CO$^{\mathrm{br}}$, O$^{\mathrm{hol}}$+CO$^{\mathrm{hol}}$, O$^{\mathrm{br}}$+CO$^{\mathrm{hol}}$ and O$^{\mathrm{hol}}$+CO$^{\mathrm{br}}$) the factor $f^{\mathrm{reac, TST}}_{i \to f} \approx 1$ in Eq.~(\ref{eq:ratereac}) is approximated by unity.  Again it is thus assumed that to a first approximation the partition functions of the initial and transition state are in the same order of magnitude, and the ratio of the two is thus close to one.

To determine the  reaction barriers, $\Delta E^{\mathrm{reac}}_{i \to f}$, the corresponding transition states are located by dragging the two reactants towards each other and mapping out the corresponding 2D-PES (O$^{\mathrm{br}}$+CO$^{\mathrm{hol}}$ and O$^{\mathrm{hol}}$+CO$^{\mathrm{br}}$). Hereby only the coordinates of the reactants along the dragging direction are kept fixed whereas all other coordinates are fully relaxed. Similar to the problem encountered when determining the diffusion barriers  the reconstructed palladium layer showed a rather high mobility with respect to a lateral shift along the substrate following the dragging direction. To circumvent this problem a two-step approach was chosen to nevertheless identify appropriate transition states. In the first step the lateral coordinates of the reconstructed palladium layer were additionally fixed and the energy is only minimized with respect to the remaining coordinates. In a second step the identified transition state geometry is relaxed with respect to the lateral coordinates of the topmost Pd layer while fixing the coordinates of the reactants O and CO, to release some of the lateral stress that was induced by fixing the lateral positions of the Pd atoms in the first step.
\begin{table}
\caption{\label{tab:reacbarrier}
Barriers for the reaction of O and CO adsorbed in different sites within the $\sqrt{5}$ surface unit cell.  All values are in eV.}
\begin{ruledtabular}
\begin{tabular}{lcccc}
\rule[0mm]{0mm}{4mm}
& O$^{\mathrm{br}}$+CO$^{\mathrm{br}}$ & O$^{\mathrm{hol}}$+CO$^{\mathrm{hol}}$ & O$^{\mathrm{br}}$+CO$^{\mathrm{hol}}$ & O$^{\mathrm{hol}}$+CO$^{\mathrm{br}}$\\[0.5ex]
\hline
\rule[0mm]{0mm}{4mm}$\Delta E^{\mathrm{reac}}$     &    1.0        &   1.6      &       0.5   &   0.9  \\[0.5ex]
\end{tabular}
\end{ruledtabular}
\end{table}
For the reaction of O$^{\mathrm{br}}$+CO$^{\mathrm{br}}$ and O$^{\mathrm{hol}}$+CO$^{\mathrm{hol}}$ this two-step approach was not necessary. Similar to the dissociative adsorption of O$_2$ an elbow plot was calculated giving the energy as a function of the O--CO distance and the height of the corresponding center of mass above the surface. Due to the symmetry of the calculated geometries no shift of the substrate layer could occur in this case. All correspondingly calculated reaction barriers are summarized in Table~\ref{tab:reacbarrier}.

For the reaction processes the energies of the initial, transition and final states are also influenced by the nearest neighbor interactions. Inspecting the geometries of the transition states they appear to be more related to the initial states and we therefore assume that the transition states are similarly affected by these interactions as the initial states.  Thus, the energy difference between the initial and transition states is not influenced by the nearest neighbor interactions and the reaction barriers are constant with respect to different configurations on the surface.

Since the reaction processes are modeled as associative desorption the corresponding time reversed process is the dissociative adsorption of CO$_2$ into the respective sites.  The adsorption barrier for such a process can be obtained from the reaction barrier and the corresponding binding energies of O and CO on the surface with respect to the free CO$_2$ molecule.  The lowest of the four resulting barriers for the dissociative adsorption of CO$_2$ (leading to CO in bridge and O in a neighboring hollow site) is still very large with $\Delta E^{\mathrm{ads}}_{\mathrm{CO}_2,\mathrm{br-hol}} = 1.4$\,eV. At a temperature of $T = 600$\,K the local sticking coefficient $\tilde{S}_{\mathrm{CO}_2,\mathrm{br-hol}}$ is even in this case then only of the order of $10^{-12}$. Even if the CO$_2$ generated at the surface is not readily transported away and a noticeable CO$_2$ pressure would build up close to the surface, the re-adsorption of CO$_2$ on the surface is therefore still negligible due to the very low  sticking coefficient and is thus not explicitly considered in the kMC simulations.

\section{Results}
Having established the setup of the kMC lattice model and the respective rate constants  for the 26 different processes based on DFT energetics and harmonic TST we proceed to the results of the kMC simulations. Although the established kMC model has a wider range of applicability, we focus here on the stability of the thin surface oxide layer during steady-state catalytic CO oxidation at the surface.

\subsection{Reproducing the thermodynamic ``phase'' diagram}
\begin{figure}
\scalebox{0.175}{\includegraphics{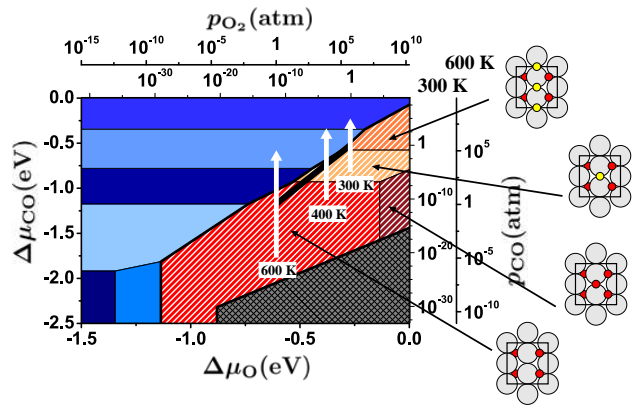}}
\caption{\label{fig4}
(Color online)
Thermodynamic surface ``phase'' diagram of the Pd(100) surface in a constrained equilibrium with and O$_2$ and CO gas-phase as described in detail in Ref. \onlinecite{rogal07}. The differently shaded areas mark the stability regions of various surface structures for a given chemical potential of the O$_2$ and CO gas-phase.
Blue/solid areas include O/CO adsorption structures on Pd(100), red/white-hatched areas comprise surface oxide structures and the grey/crosshatched area represents the stability range of PdO bulk oxide.
For convenience the dependence of the chemical potential of the two gas-phases is translated into pressure scales for $T=300$\,K and $T=600$\,K (upper $x$ axes and right $y$ axes).
For the stability region of the surface oxide structures additionally schematic figures of the different surface configurations are shown.  Grey, large spheres represent Pd atoms, red (dark) small spheres oxygen atoms and yellow (light) small spheres CO molecules.
The white arrows indicate the pressure conditions of the kMC simulations for a temperature of $T=300$\,K, $T=400$\,K and $T=600$\,K.  The oxygen pressure is always fixed at $p_{\mathrm{O}_2}=1$\,atm, while the CO pressure is varied between $10^{-5} \leq p_{\mathrm{CO}} \leq 10^{5}$\,atm.
}
\end{figure}

In order to make contact to our preceding constrained thermodynamics work~\cite{rogal07},
we perform the kMC simulations in a first step  only considering adsorption, desorption and diffusion processes, whereas the reaction processes are excluded.  The occupation of the different sites at the surface under steady-state conditions reflects then the same situation as a surface in a constrained thermodynamic equilibrium with an oxygen and CO gas-phase, but including the effect of configurational entropy~\cite{reuter03}.
The results of the constrained thermodynamics approach are summarized in Fig.~\ref{fig4} (for a detailed discussion see Ref.~\onlinecite{rogal07}).  In the shown ``phase'' diagram the most stable surface structures for any given chemical potential of the O$_2$ and CO gas-phase are presented (the respective pressure scales for $T=300$\,K and 600\,K are given in the upper $x$-axes and right $y$-axes).
The stability regions of the different surface structures are denoted by differently shaded areas.  For the here discussed kMC simulations it is particularly important that the different ``phases'' can be subdivided into three groups.  All areas colored in blue/solid represent structures with O and CO adsorbed on the pristine Pd(100) surface, all red/white-hatched areas comprise structures involving the reconstructed $\sqrt{5}$ surface oxide, and the grey/crosshatched area marks the stability region of the bulk oxide.  As discussed in Section~\ref{subsec:kmcmodel} the here chosen kMC lattice model is based on the lattice structure of the $\sqrt{5}$ surface oxide and does therefore not allow for changes between this structure and the Pd(100) surface or the bulk oxide.  Consequently, the constrained kMC simulations can only reproduce the part of the ``phase'' diagram where structures based on the surface oxide are most stable, i.e. the red/white-hatched region in Fig.~\ref{fig4}.

Within the constrained kMC simulations we now evaluate the steady-state average occupation of bridge and hollow sites and compare the obtained structures to the structures predicted within the thermodynamic surface ``phase'' diagram.  As mentioned above the two approaches should yield equivalent results, except at the ``phase'' boundaries where configurational entropy is expected to become important~\cite{reuter03}.
The simulations are performed for $(T,p)$-conditions that cover the entire stability region of the surface oxide (red/white-hatched region in Fig.~\ref{fig4}).  Specifically, this means for $T=300$\,K pressures of $p_{\mathrm{O}_2} = 10^{-10} - 10^{10}$\,atm and $p_{\mathrm{CO}} = 10^{-10} - 1$\,atm, and for $T=600$\,K pressures of $p_{\mathrm{O}_2} = 1 - 10^{10}$\,atm and $p_{\mathrm{CO}} = 1 - 10^{6}$\,atm, respectively.
We find that the kMC simulations reproduce accurately all four different surface oxide ``phases'' that appear in this part of the constrained ``phase'' diagram.  In all four ``phases'' all hollow sites are always filled with oxygen.  The bridge sites are either empty, half filled with O  or CO, or completely filled with CO  depending on the respective gas-phase conditions and as detailed by the schematic representations in Fig.~\ref{fig4}. As expected from the common underlying DFT energetics, we therefore find the results of the constrained kMC simulations to be fully consistent with those obtained previously within the constrained thermodynamics approach.  The more extended sampling of configurations due to the LGH expansion within the kMC simulations does not lead to the appearance of new, stable ordered structures in the constrained ``phase'' diagram and thus confirms that the relevant configurations were included in the preceding thermodynamic approach. The effect of configurational entropy in the constrained kMC simulations is thus verified to ``only'' smear out the transitions between the different ordered ``phases'' at the finite temperatures considered~\cite{reuter03}, but does not affect the location of the ``phase'' boundaries themselves.  Due to the hereby confirmed consistency between the two approaches, we can directly compare the results in the next Section, when allowing the reaction events to occur in the kMC simulation. This brings us in the position to directly address the effect of the kinetics of the ongoing chemical reactions on the surface structure and composition.

\subsection{Onset of surface oxide decomposition under reaction conditions}
By allowing the four different reaction processes to occur in the kMC simulations, we now proceed to evaluate the stability of the surface oxide structure under reaction conditions. Since the kMC lattice is fixed to the structure of the $\sqrt{5}$ surface oxide, the explicit structural change corresponding to the transition from the surface oxide to a pristine Pd(100) surface as predicted in the thermodynamic ``phase'' diagram (red/white-hatched areas $\to$ blue solid areas in Fig.~\ref{fig4}) cannot directly be addressed. Instead, we start our simulations always in a $(T,p)$-range where the surface oxide is definitely the most stable ``phase'' (high oxygen and low CO chemical potential), and where our kMC lattice model is certainly valid.  Keeping the oxygen pressure and the temperature fixed we then increase the CO pressure as indicated by the white, vertical arrows in Fig.~\ref{fig4}.  Exceeding a certain pressure ratio of $p_{\mathrm{O}_2}/p_{\mathrm{CO}}$,
we expect the onset of the decomposition of the $\sqrt{5}$ structure to be characterized by a depletion of the otherwise always essentially fully with oxygen occupied hollow sites.
In the kMC simulations we therefore monitor
the average occupation of hollow sites by oxygen atoms, $\overline{\theta}_{\mathrm{O}^{\mathrm{hol}}}$, during steady-state.  If all hollow sites are occupied by oxygen, i.e. $\overline{\theta}_{\mathrm{O}^{\mathrm{hol}}} \approx 100\%$, then the overall surface composition clearly represents the intact surface oxide structure (this is the case in all four ``phases'' of the surface oxide structure in Fig.~\ref{fig4}).
If on the other hand at increasing $p_{\mathrm{CO}}$ oxygen atoms get increasingly depleted beyond the amount of thermal fluctuations that create isolated, uncorrelated vacancies or double vacancies due to oxidation reaction or oxygen desorption events, then the deficiency of O atoms at the surface might destabilize the $\sqrt{5}$ surface structure and lead to a local lifting of the surface oxide reconstruction.
Here, we therefore interpret a drop of the average occupation below 90\% as the onset of a destabilization of the surface oxide.
Since the average occupation is a global measure over the whole simulation cell this quantity would be a bad indicator, if there was an appreciable effective attraction between formed O vacancies.  A decrease in the average occupation of hollow sites by oxygen of 10\% could then involve a complete, local depletion of 10\% of the surface oxide area.  In our kMC simulations such a locally concentrated depletion of oxygen atoms is not observed.
Instead, we find the formed oxygen vacancies to be distributed over the entire simulation cell and are thus confident that monitoring the average occupation, $\overline{\theta}_{\mathrm{O}^{\mathrm{hol}}}$, is a suitable measure to determine the onset of surface oxide decomposition.

\begin{figure}
\scalebox{0.6}{\includegraphics{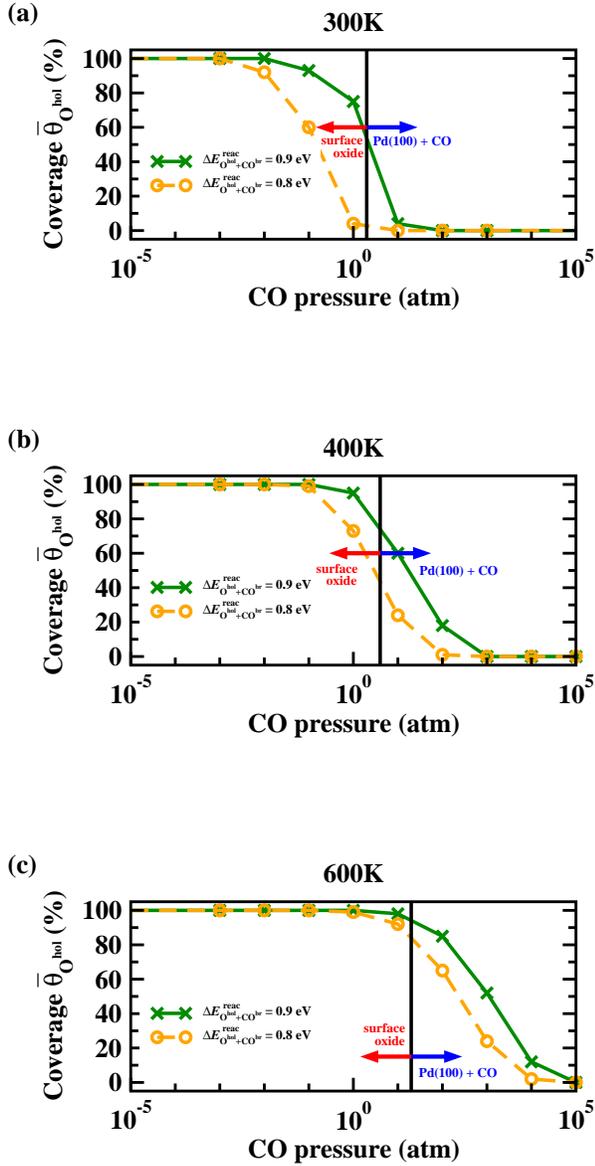}}
\caption{\label{fig5}
(Color online)
Average occupation of hollow sites by oxygen vs. CO pressure for $T = 300$\,K, $T = 400$\,K and $T = 600$\,K.  The vertical black line marks the boundary between the surface oxide and a CO covered Pd(100) surface as determined within the constrained thermodynamics approach. The influence of the barrier for the O$^{\mathrm{hol}} +$CO$^{\mathrm{br}}$ reaction is illustrated by showing the kMC simulation results for two barrier values, 0.9\,eV and 0.8\,eV (see text).}
\end{figure}

The simulations are performed for three different temperatures, $T=300, 400$ and 600\,K.  For each temperature the  oxygen pressure is fixed to $p_{\mathrm{O}_2} = 1$\,atm and the simulations are run for different CO pressures between  $p_{\mathrm{CO}} = 10^{-5} - 10^5$\,atm  covering the gas-phase conditions marked by the white arrows in Fig.~\ref{fig4}.
In Fig.~\ref{fig5} the corresponding simulation results are summarized.
The steady-state average occupation of hollow sites with oxygen, $\overline{\theta}_{\mathrm{O}^{\mathrm{hol}}}$, is plotted vs. the CO pressure.  The solid (green) lines represent the results obtained by using the calculated reaction barriers as listed in Table~\ref{tab:reacbarrier}.
As has been discussed in Section~\ref{sec:theory}, several, partly crude approximations were employed in the determination of the individual rate constants entering the kMC simulations.  Before we proceed to analyze the obtained simulation results, we therefore critically assess the propagation of these underlying approximations and uncertainties to the final kMC results.  We checked on this error propagation by systematically varying the rate constants over several orders of magnitude (corresponding to variations in the respective energetic barriers of $\pm 0.2$\,eV) and each time monitoring the effect on the average occupation displayed in Fig.~\ref{fig5}.  In almost all cases, variations of the rate constants had an insignificant effect, which as already described concerns in particular all diffusion processes. Only if altering the barrier for the reaction process of an oxygen in  a hollow site with a CO in a neighboring bridge site a noticeable change in the average surface populations occurs. To get an estimate for the corresponding uncertainties in our presented results, we therefore include
in Fig.~\ref{fig5}  additionally the results for simulations using a 0.1\,eV lower barrier for this reaction process O$^{\mathrm{hol}} +$CO$^{\mathrm{br}}$, i.e.  $\Delta E^{\mathrm{reac}}_{\mathrm{O}^{\mathrm{hol}}+\mathrm{CO}^{\mathrm{br}}} = 0.8$\,eV, while all other barriers have been left unchanged (dashed (orange) lines). The differences between these results (dashed (orange) line) and the results for the unmodified barrier (solid (green) line) provide then an indication of the uncertainty underlying our approach.

Discussing first the top graph in Fig.~\ref{fig5} it can be seen that for a temperature of $T = 300$\,K and an oxygen pressure of $p_{\mathrm{O}_2} = 1$\,atm the surface oxide is clearly stabilized for CO pressures up to $p_{\mathrm{CO}} < 10^{-1}$\,atm, i.e. under such gas-phase conditions $\overline{\theta}_{\mathrm{O}^{\mathrm{hol}}} > 95\,\%$.
If the CO pressure is further increased the O population at  hollow sites starts to decrease and for CO pressures of $p_{\mathrm{CO}} > 1$\,atm the surface oxide is certainly completely destabilized ($\overline{\theta}_{\mathrm{O}^{\mathrm{hol}}} \approx 0\,\%$).  For the slightly lower reaction barrier (dashed (orange) line) a steeper decrease in the oxygen population is found, and the surface oxide is stabilized for CO pressures of $p_{\mathrm{CO}} < 10^{-2}$\,atm.
To compare these results to the ones obtained within the constrained thermodynamics approach the stability region of the surface oxide as determined in the ``phase'' diagram in Fig.~\ref{fig4} is indicated by the vertical black lines in the three graphs of Fig.~\ref{fig5}.  Following the white arrow in the ``phase'' diagram in Fig.~\ref{fig4} for $T = 300$\,K and $p_{\mathrm{O}_2} = 1$\,atm  this boundary between the surface oxide structure and a CO covered Pd(100) surface is reached for a CO pressure of $p_{\mathrm{CO}} \approx 2$\,atm. Within the constrained thermodynamics approach the surface oxide will thus be stable for $p_{\mathrm{CO}} \lesssim 2$\,atm, whereas for $p_{\mathrm{CO}} \gtrsim 2$\,atm the CO covered Pd(100) surface is the most stable ``phase''. Comparing this to the kMC simulations which include the effect of the CO$_2$ formation kinetics, we see that the depletion of oxygen atoms in hollow sites and thus the destabilization of the surface oxide structure starts already at slightly lower CO pressures compared to the transition from the stability region of the surface oxide to the one of a CO covered Pd(100) surface in the ``phase'' diagram. Although we find the kMC simulations to be dominated by adsorption/desorption processes of the reactants (as anticipated in the constrained thermodynamic approach assuming equilibrium with the gas-phase reservoirs), the kinetics of the ongoing oxidation reaction thus still leads to a slight decrease in the stability of the surface oxide. In turn the kMC simulations predict that at a temperature of $T=300$\,K oxygen-rich conditions ($p_{\mathrm{CO}}/p_{\mathrm{O}_2} \approx 1:10$) are needed to stabilize the surface oxide structure.

In Fig.~\ref{fig5}(b) corresponding results are shown for $T=400$\,K.  Compared to the simulations at $T=300$\,K much more reaction processes take place in the kMC simulations at this higher temperature.  For CO pressures of $p_{\mathrm{CO}} > 10$\,atm though almost no reaction processes are detected in the steady-state, since under these conditions the surface is already completely filled with CO. Then again this corresponds to gas-phase conditions where also in the constrained thermodynamics approach a CO covered Pd(100) surface is the most stable structure (cf. Fig.~\ref{fig4}).  Similar to the results for $T=300$\,K the surface oxide structure is completely stabilized for CO pressures of $p_{\mathrm{CO}} < 10^{-1}$\,atm. At a pressure ratio of $p_{\mathrm{CO}}/p_{\mathrm{O}_2} = 2:1$, i.e. here $p_{\mathrm{CO}} = 2$\,atm, there are still 94\,\% of all hollow sites occupied by oxygen (solid (green) line in Fig.~\ref{fig5}(b)). This result depends, however, critically on the barrier for the reaction of O$^{\mathrm{hol}}$+CO$^{\mathrm{br}}$.  If this barrier is decreased by only 0.1\,eV the occupation drops to 70\,\%, and at an even lower barrier of $\Delta E^{\mathrm{reac}}_{\mathrm{O}^{\mathrm{hol}}+\mathrm{CO}^{\mathrm{br}}} = 0.7$\,eV  $\overline{\theta}_{\mathrm{O}^{\mathrm{hol}}}$ decreases even further to $\sim 30$\,\% under these conditions.
Whether or not the surface oxide prevails at this temperature for the catalytically most relevant stoichiometric partial pressure ratio can thus not be safely concluded in light of the uncertainties of our approach.
For CO pressures lower than $p_{\mathrm{CO}} < 10^{-1}$\,atm, however, the surface oxide is definitely stabilized,
just as much as it is definitely destabilized at CO pressures of $p_{\mathrm{CO}} > 10$\,atm. We can therefore determine
the onset of the surface oxide decomposition to fall into the pressure range 0.1\,atm $< p_{\mathrm{CO}} < 10$\,atm.  Even considering the uncertainties in the reaction rate constants and the limitation to a single kMC lattice,  these results agree thus rather nicely with the reactor STM experiments by Hendriksen \emph{et al.}~\cite{hendriksen04} which have been performed under similar temperature and pressure conditions ($T=408$\,K, $p_{\mathrm{tot}} = p_{\mathrm{CO}} + p_{\mathrm{O}_2} = 1.23$\,atm).  Depending on the pressure ratio of oxygen and CO Hendriksen \emph{et al.} observed a change in the morphology of the surface, which was assigned to a change from an oxidic to a reduced surface structure and oxygen-rich feeds were required to stabilize the oxidic structure.

The simulation results for the highest considered temperature of $T=600$\,K are shown in Fig.~\ref{fig5}(c).  We find that for this elevated temperature the surface oxide is now actually stable up a rather sizable CO pressures.  Only for CO pressures as high as $p_{\mathrm{CO}} = 10$\,atm the surface oxide starts to decompose ($\overline{\theta}_{\mathrm{O}^{\mathrm{hol}}} < 95\,\%$), which already coincides almost with  the border of its stability as determined in the constrained thermodynamics approach (black vertical line).  In the ``phase'' diagram in Fig.~\ref{fig4} it can be seen that for a temperature of $T=600$\,K and an oxygen pressure of $p_{\mathrm{O}_2} = 1$\,atm CO is not stabilized in large amounts as adsorbate on the surface oxide.  The CO is thus not readily available as adsorbed reaction partner and the stability of the surface oxide is hardly affected by the ongoing catalytic CO$_2$ formation. Up to the stoichiometric pressure ratio of $p_{\mathrm{CO}}/p_{\mathrm{O}_2} = 2:1$ no decomposition of the surface oxide structure occurs, and as apparent from Fig.~\ref{fig5}(c) this result holds even in light of the estimated uncertainty of the underlying DFT energetics. Comparing the critical $p_{\rm CO}/p_{\rm O_2}$ ratio determined for the decomposition onset in the temperature range $T = 300 - 600$\,K, we can thus clearly identify an increasing stability of the surface oxide with increasing temperature, which at the highest temperatures studied reaches well up to the catalytically most relevant feeds. Furthermore we find that under these feeds the simulated turnover frequencies for the intact surface oxide alone are already of similar order of magnitude as those reported by Szanyi and Goodman~\cite{szanyi94} for the Pd(100) surface under comparable gas-phase conditions. While a quantitative comparison is outside the scope of the present work, we note that contrary to the prevalent general preconception at least this particular surface oxide is thus clearly not ``inactive'' with respect to the oxidation of CO.

\section{Summary}

We have employed first-principles kMC simulations to investigate the stability of the $\sqrt{5}$ surface oxide structure at Pd(100) against CO-induced decomposition under steady-state operation conditions of the CO oxidation reaction. The focus has been particularly set on the thin surface oxide, since this structure was identified as the most stable oxidic ``phase'' under catalytically relevant gas-phase conditions in a preceding constrained first-principles thermodynamics study~\cite{rogal07}. The developed kMC model describes the CO oxidation on the $\sqrt{5}$ surface oxide, accounting for all non-concerted processes that are possible on the chosen kMC lattice. This includes 26 site and element specific adsorption, desorption, diffusion and reaction processes.  The respective kMC rate constants have been obtained from DFT total energy calculations together with harmonic transition state theory.

Monitoring the onset of the surface oxide decomposition at increasing CO pressures, the central result of the kMC simulations is that much in contrast to thicker bulk-like oxide films the stability of the $\sqrt{5}$ surface oxide can well extend to gas-phase conditions that are relevant for technological CO oxidation catalysis. While the accuracy of present-day DFT functionals does not permit us to conclude on precise values for the critical CO:O$_2$ partial pressure ratio above which decomposition sets in, our results show undoubtedly that at ambient pressures this ratio is close to the most interesting stoichiometric feed conditions ($p_{\rm CO}:p_{\rm O_2} = 2:1$) and will furthermore shift decisively towards more CO-rich conditions with increasing temperature.

On the basis of these results we conclude that the surface oxide structure can actually be present under catalytic reaction conditions at ambient pressures and our simulated turnover frequencies indicate that it is then also catalytically active towards the oxidation of CO. For the catalytically most relevant feeds the system appears furthermore to be rather close to a transition between the $\sqrt{5}$ surface oxide structure and a reactant covered Pd(100) surface. Particularly in stoichiometric to slightly O-rich feeds even local fluctuations in both temperature or the reactant partial pressures might then lead to significant local or global changes in the structure and composition of the surface. Under steady-state operation conditions oscillations between these two states are thus conceivable, so that in its active state the catalyst surface might exhibit parts that correspond to Pd(100) covered by the reactants, as well as by patches of surface oxide, and potentially it is precisely the on-going formation and decomposition of the oxidic structure that is key to understand the catalytic function of this surface under such technologically relevant pressures. For a full understanding of the catalytic CO oxidation at the Pd(100) surface in corresponding environments it thus appears not to be sufficient to concentrate on either the reduced Pd(100) surface or on an oxidic structure, but both surface states as well as the morphological transition between the two will have to be included in an appropriate model.

\section{Acknowledgment}
The EU is acknowledged for financial support under contract NMP3-CT-2003-505670 (NANO$_2$), and the DFG for support within the priority program SPP1091.


\begin{thebibliography}{99}


\bibitem{address}
Present address: Van 't Hoff Institute for Molecular Sciences, University of Amsterdam,
Nieuwe Achtergracht 166, 1018 WV Amsterdam, The Netherlands.

\bibitem{peden86}
C.H.F. Peden and D.W. Goodman, J. Phys. Chem {\bf 90}, 1360 (1986).

\bibitem{over03}
H. Over and M. Muhler,
Prog. Surf. Sci. {\bf 72}, 3 (2003).

\bibitem{reuter06a}
K. Reuter,
Oil \& Gas Science and Technology  Rev. IFP \textbf{61}, 471 (2006);
http://ogst.ifp.fr

\bibitem{reuter04}
K. Reuter and M. Scheffler,
Appl. Phys. A {\bf 78}, 793 (2004).

\bibitem{reuter03}
K. Reuter and M. Scheffler,
Phys. Rev. Lett. {\bf 90}, 046103 (2003); Phys. Rev. B \textbf{68}, 045407 (2003).

\bibitem{rogal07}
J. Rogal, K. Reuter, and M. Scheffler,
Phys. Rev. B \textbf{75}, 205433 (2007).

\bibitem{orent82}
T. W. Orent, and S. D. Bader,
Surf. Sci. \textbf{115}, 323 (1982).

\bibitem{chang88a}
S.-L. Chang, and P. A. Thiel,
J. Chem. Phys. \textbf{88}, 2071 (1988).

\bibitem{chang88b}
S.-L. Chang, P. A. Thiel, and J. W. Evans,
Surf. Sci. \textbf{205}, 117 (1988).

\bibitem{zheng02}
G. Zheng, and E. I. Altman,
Surf. Sci. \textbf{504}, 253 (2002).

\bibitem{lundgren04}
E. Lundgren,
J. Gustafson, A. Mikkelsen, J.N. Andersen, A. Stierle, H. Dosch,
M. Todorova, J. Rogal, K. Reuter, and M. Scheffler,
Phys. Rev. Lett. \textbf{92}, 046101 (2004).

\bibitem{todorova03}
M. Todorova,
E. Lundgren, V. Blum, A. Mikkelsen, S. Gray, J. Gustafson,
Borg, J. Rogal, K. Reuter, J.N. Andersen, and M. Scheffler,
Surf. Sci. \textbf{541}, 101 (2003).

\bibitem{kostelnik07}
P. Kosteln\'{\i}k, N. Seriani, G. Kresse, A. Mikkelsen,
E Lundgren, V. Blum, T. \v{S}ikola, P. Varga, and M. Schmid,
Surf. Sci. \textbf{601}, 1574 (2007).

\bibitem{rogal07a}
J. Rogal, K. Reuter, and M. Scheffler,
Phys. Rev. Lett. \textbf{98}, 046101 (2007).

\bibitem{hendriksen04}
B.L.M. Hendriksen, S.C. Bobaru, and J.W.M. Frenken,
Surf. Sci. {\bf 552}, 229 (2004).

\bibitem{bortz75}
A. B. Bortz, M. H. Kalos, and J. L. Lebowitz,
J. Comp. Phys. \textbf{17}, 10 (1975).

\bibitem{gillespie76}
D. T. Gillespie,
J. Comp. Phys. \textbf{22}, 403 (1976).

\bibitem{voter86}
A. F. Voter,
Phys. Rev. B \textbf{34}, 6819 (1986).

\bibitem{kang89}
H. C. Kang, and W. H. Weinberg,
J. Chem. Phys. \textbf{90}, 2824 (1989).

\bibitem{fichthorn91}
K. A. Fichthorn, and W. H. Weinberg,
J. Chem. Phys. \textbf{95}, 1090 (1991).

\bibitem{ruggerone97}
P. Ruggerone, C. Ratsch, and M. Scheffler,
``Density-functional theory of epitaxial growth of metals,''
in The Chemical Physics of Solid Surfaces Vol. 8:
Growth and Properties of Ultrathin Epitaxial Layers,
edited by D. A. King and D. P. Woodruff
(Elsevier Science, Amsterdam, 1997)

\bibitem{ovesson99}
S. Ovesson, A. Bogicevic, and B. I. Lundqvist,
Phys. Rev. Lett. \textbf{83}, 2608 (1999).

\bibitem{hansen00a}
E. W. Hansen and M. Neurock,
J. Catal. \textbf{196}, 241 (2000).

\bibitem{hansen00b}
E. W. Hansen and M. Neurock,
Surf. Sci. \textbf{464}, 91 (2000).

\bibitem{fichthorn00}
K. A. Fichthorn and M. Scheffler,
Phys. Rev. Lett. \textbf{84}, 5371 (2000).

\bibitem{kratzer02}
P. Kratzer and M. Scheffler,
Phys. Rev. Lett. \textbf{88}, 036102 (2002).

\bibitem{reuter04b}
K. Reuter, D. Frenkel, and M. Scheffler,
Phys. Rev. Lett {\bf 93}, 116105 (2004).

\bibitem{reuter06}
K. Reuter and M. Scheffler,
Phys. Rev. B {\bf 73}, 045433 (2006).

\bibitem{gross98}
A. Gross,
Surf. Sci. Rep. \textbf{32}, 293 (1998).

\bibitem{sjoestedt00}
E. Sj\"ostedt, L. Nordstr\"om, and D. J. Singh,
Solid State Comm. \textbf{114}, 15 (2000).

\bibitem{madsen01}
G. K. H. Madsen, P. Blaha, K. Schwarz, E. Sj\"ostedt, and L. Nordstr\"om,
Phys. Rev. B \textbf{64}, 195134 (2001).

\bibitem{wien2k}
P. Blaha, K. Schwarz, G.K. Madsen, D. Kvasnicka, and J. Luitz,
\textsf{WIEN2k}, Techn. Universit\"at Wien, Austria (2001).
ISBN 3-9501031-1-2.

\bibitem{perdew96}
J.P. Perdew, K. Burke, and M. Ernzerhof,
Phys. Rev. Lett. {\bf 77}, 3865 (1996).

\bibitem{rogal06}
J. Rogal and K. Reuter,
``Ab initio Atomistic Thermodynamics for Surfaces: A Primer'' in:
``Experiment, Modeling and Simulation of Gas-Surface Interactions for Reactive Flows in
Hypersonic Flights'', p 2-1 � 2-18,
Educational Notes RTO-EN-AVT-142,
Neuilly-sur-Seine (2007).

\bibitem{stampfl99}
C. Stampfl, H.J. Kreuzer, S.H. Payne, H. Pfn\"ur, and M. Scheffler,
Phys. Rev. Lett. \textbf{83}, 2993 (1999).

\bibitem{reuter05}
K. Reuter, C. Stampfl, and M. Scheffler,
``Ab initio atomistic thermodynamics and statistical mechanics of surface properties and functions,'' in
Handbook of Materials Modeling, Vol. 1,
edited by Sidney Yip
(Springer Berlin Heidelberg, 2005, 149-194)

\bibitem{fontaine94}
D. De Fontaine,
in Statistics and dynamics of alloy phase transformations,
NATO ASI Series, Plenum Press, New York, 1994.

\bibitem{sanchez84}
J. M. Sanchez, F. Ducastelle, and D. Gratias,
Physica A \textbf{128}, 334 (1984).

\bibitem{zunger94}
A. Zunger,
``First principles statistical mechanics of semiconductor alloys and intermetallic
compounds'', in Statistics and dynamics of alloy phase transformations,
NATO ASI Series, Plenum Press, New York, 1994.

\bibitem{kiejna06}
A. Kiejna, G. Kresse, J. Rogal, A. De Sarkar, K. Reuter, and M. Scheffler,
Phys. Rev. B \textbf{73}, 035404 (2006).

\bibitem{szanyi94}
J. Szanyi and D.W. Goodman, 
J. Phys. Chem. \textbf{98}, 2972 (1994).







\end{thebibliography}
\end{document}